\documentclass[10pt,conference,letterpaper]{IEEEtran}

\usepackage{layouts}
\usepackage[T1]{fontenc}
\usepackage[utf8]{inputenc}
\usepackage[english]{babel}

\usepackage{ifthen}

\usepackage{times}
\usepackage{balance}
\usepackage{csquotes}
\usepackage{microtype}

\usepackage{xcolor}
\usepackage{graphicx}

\usepackage{bm}
\usepackage{amsmath}
\DeclareMathOperator{\sgn}{sgn}
\usepackage{amsthm}
\usepackage{amsfonts}
\usepackage{amssymb}

\usepackage{siunitx}

\usepackage[
    bibstyle=ieee,
    citestyle=numeric,
    minnames=1,
    maxnames=6,
    mincitenames=1,
    maxcitenames=1,
    autocite=inline,
    sortcites=true,
    dashed=false,
    labelnumber=true]{biblatex} 
\usepackage[acronym,hyperfirst=false,nohypertypes={acronym}]{glossaries}

\usepackage[font=small,belowskip=-1.1em]{caption}

\usepackage{tabu}

\newcolumntype{C}{>{\centering\arraybackslash}X} 
\usepackage{booktabs}
\usepackage{xspace}

\usepackage[disable]{todonotes}

\usepackage{url}
\usepackage[colorlinks=false,hidelinks=true]{hyperref}
\usepackage{cleveref}
\usepackage[shortcuts]{extdash}

\makeatletter
\newcounter{IEEE@bibentries}
\renewcommand\IEEEtriggeratref[1]{%
  \renewbibmacro{finentry}{%
    \stepcounter{IEEE@bibentries}%
    \ifthenelse{\equal{\value{IEEE@bibentries}}{#1}}
    {\finentry\@IEEEtriggercmd}
    {\finentry}%
  }%
}
\makeatother

\DeclareCaptionLabelSeparator{iaria}{.\quad}
\graphicspath{{./figures/}}
\DeclareGraphicsExtensions{.pdf,.jpeg,.png}

\crefname{figure}{Fig.}{Fig.}
\crefname{table}{Table}{Tables}
\crefname{section}{Section}{Sections}

\newacronym[shortplural={CPS}]{CPS}{CPS}{cyber-physical system}
\newacronym{AI}{AI}{Artificial Intelligence}
\newacronym{AL}{AL}{Adversarial Learning}
\newacronym{ANN}{ANN}{Artificial Neural Network}
\newacronym{ARL}{ARL}{Adversarial Resilience Learning}
\newacronym{GAN}{GAN}{Generative Adversarial Network}
\newacronym{GPGPU}{GPGPU}{General-Purpose Graphics Processing Unit}
\newacronym{DL}{DL}{Deep Learning}
\newacronym{GRU}{GRU}{Gated Recurrent Unit}
\newacronym{ICT}{ICT}{Information and Communication Technology}
\newacronym{LSTM}{LSTM}{Long-Short Term Memory}
\newacronym{AEG}{AEG}{Attack Execution Graph}
\newacronym{ML}{ML}{Machine Learning}
\newacronym{PoC}{PoC}{Proof-of-Concept}
\newacronym{RL}{RL}{Reinforcement Learning}
\newacronym{DNN}{DNN}{Deep Neural Network}
\newacronym{RNN}{RNN}{Recurrent Neural Network}
\newacronym{CNN}{CNN}{Convolutional Neural Network}
\newacronym{SIMD}{SIMD}{Single Instruction, Multiple Data}
\newacronym{DoE}{DoE}{design of experiments}
\newacronym{DNC}{DNC}{Differentiable Neural Computer}
\newacronym{PV}{PV}{Photovoltaic}
\newacronym{MV}{MV}{medium voltage}
\newacronym{CHP}{CHP}{combined heat and power}
\newacronym{STL}{STL}{Signal Temporal Logic}
\newacronym{MTL}{MTL}{Metric Temporal Logic}
\newacronym{MITL}{MITL}{Metric Interval Temporal Logic}
\newacronym{AV}{AV}{Autonomous Vehicle}
\newacronym[shortplural={MAS}]{MAS}{MAS}{Multi Agent System}
\newacronym{RTU}{RTU}{Remote Terminal Unit}
\newacronym{TCP}{TCP}{Transmission Control Protocol}
\newacronym{LPEP}{LPEP}{Lightweight Power Exchange Protocol}
\newacronym{FMI}{FMI}{Functional Mock-Up Interface}
\newacronym{XML}{XML}{eXtensible Markup Language}
\newacronym{IoT}{IoT}{Internet of Things}
\newacronym{SIEM}{SIEM}{Security Information and Event Management}
\newacronym{APT}{APT}{Advanced Persistent Threads}
\newacronym{TAL}{TAL}{Thread Agents Library}
\newacronym{CBN}{CBN}{Causal Bayesian Network}
\newacronym{DER}{DER}{distributed energy resource}
\newacronym{CIGRE}{CIGRE}{International Council on Large Electric Systems}
\newacronym{IPF}{IPF}{interconnection power flow}
\newacronym{PDF}{PDF}{probability density function}
\newacronym{TSO}{TSO}{transmission system operator}
\newacronym{DSO}{DSO}{distribution system operator}
\newacronym{TG}{TG}{transmission grid}
\newacronym{DG}{DG}{distribution grid}
\newacronym{ADG}{ADG}{active distribution grid}
\newacronym{OLTC}{OLTC}{on-load tap changer}
\newacronym{FOR}{FOR}{feasible operation region}
\newacronym{OPF}{OPF}{optimal power flow}
\newacronym{PSO}{PSO}{particle swarm optimization}
\newacronym{COHDA}{COHDA}{Combinatorial Optimization
Heuristic for Distributed Agents}

\newacronym{PYRATE}{PYRATE}{Polymorphic agents as cross-sectional software technology for the analysis of the operational safety of cyber-physical systems}

\newglossaryentry{resilience}{%
  name=resilience,
  description={The ability of a system to withstand unforseen, rare and
  potentially catastrophic events and to self-heal or improve in
  reaction to these events. Ideally resilience is increasing monotonously.}
}

\begin{document}
\title{\Large\bfseries%
    Pointing out the Convolution Problem of Stochastic Aggregation Methods for the
    Determination of Flexibility Potentials at Vertical System Interconnections}
\author{\IEEEauthorblockN{Johannes Gerster\\ and Sebastian Lehnhoff}
\IEEEauthorblockA{Dept. of Computing Science\\
CvO Universität Oldenburg\\
\texttt{johannes.gerster@uol.de}}
\and
\IEEEauthorblockN{Marcel Sarstedt\\ and Lutz Hofmann}
\IEEEauthorblockA{Inst. for Electric Power Systems\\
Leibniz Universität Hannover\\
\texttt{sarstedt@ifes.uni-hannover.de}}
\and
\IEEEauthorblockN{Eric MSP Veith}
\IEEEauthorblockA{%
    OFFIS e.V.\\
    R\&D Division Energy\\
    Oldenburg, Germany\\
    \texttt{eric.veith@offis.de}}}

\maketitle

\begin{abstract}
The increase of generation capacity in the area of responsibility of the
\gls{DSO} requires strengthening of coordination between \gls{TSO} and
\gls{DSO} in order to prevent conflicting or counteracting use of flexibility
options. For this purpose, methods for the standardized description and
identification of the aggregated flexibility potential of \glspl{DG} are
developed. Approaches for identifying the \gls{FOR} of \glspl{DG} can be
categorized into two main classes: Data-driven/stochastic approaches and
optimization based approaches.  While the latter have the advantage of working
in real-world scenarios where no full grid models exist, when relying on
nai\"ve sampling strategies, they suffer from poor coverage of the edges of the
\gls{FOR}. 
To underpin the need for improved sampling strategies for data-driven
approaches, in this paper we point out and analyse the shortcomings of nai\"ve
sampling strategies with focus on the problem of leptocurtic distribution of
resulting \glspl{IPF}.  We refer to this problem as \emph{convolution problem},
as it can be traced back to the fact that the \gls{PDF} of the sum of two or
more independent random variables is the convolution of their respective
\glspl{PDF}.  To demonstrate the \emph{convolution problem}, we construct
a series of synthetic 0.4~kV feeders, which are characterized by an increasing
number of nodes and apply a sampling strategy to them that draws set-values for
the controllable \glspl{DER} from independent uniform distributions.  By
calculating the power flow for each sample in each feeder, we end up with
a collapsing \gls{IPF} point cloud clearly indicating the \emph{convolution
problem}.
\end{abstract}


\section{Introduction}
\label{sec:introduction}
The increasing share of \glspl{DER} in the electrical energy system leads to
new challenges for both, \gls{TSO} and \gls{DSO}. Flexibility services for
congestion management and balancing, so far mostly provided by large scale
thermal power plants directly connected to the \gls{TG}, increasingly have to
be provided by \glspl{DER} connected to the \gls{DG}.  Thus, \glspl{DG} evolve
from formerly mostly passive systems to \glspl{ADG} that contain a variety of
controllable components interconnected via communication infrastructure and
whose dynamic behaviour is characterized by higher variability of power flows
and greater simultaneity factors.

\gls{TSO}-\gls{DSO} coordination is an important topic which has been pushed by
ENTSO\=/E during the last years~\cite{entso-e2015, entso-e2015a, entso-e2017,
entso-e2019}. Coordination between grid operators has to be strengthened to
prevent conflicting or counteracting use of flexibility
options~\cite{sarstedt2019}. To reduce complexity for \glspl{TSO} at the
\gls{TSO}/\gls{DSO} interface and to enable \glspl{TSO} to consider the
flexibility potential of \glspl{DG} in its operational management and
optimization processes, methods are needed which allow for the determination
and standardized representation of the aggregated flexibility potential of
\glspl{DG}.

The aggregated flexibility potential of a \gls{DG} can be described as region
in the PQ-plane that is made up from the set of feasible
\glspl{IPF}~\cite{mayorgagonzalez2018}. Thereby, feasible \glspl{IPF} are
\glspl{IPF} which can be realized by using the flexibilities of controllable
\glspl{DER} and controllable grid components such as \gls{OLTC} transformers in
compliance with grid constraints i.e.,\ voltage limits and maximum line
currents.

In the literature, there are various concepts to determine the \gls{FOR} of
\glspl{DG}. They can be categorized into two main classes:
Data-driven/stochastic approaches and optimization-based approaches.

For data-driven approaches, the general procedure is such that a set of random
control scenarios is generated by assigning set-values from a uniform
distribution to each controllable unit. By means of load flow calculations the
resulting \glspl{IPF} are determined for each control scenario and classified
into feasible \glspl{IPF} (no grid constraints are violated) and non-feasible
\glspl{IPF} (at least one grid constraint is violated). The resulting point
cloud of feasible \glspl{IPF} in the PQ-plane serves as stencil for the
\gls{FOR}~\cite{mayorgagonzalez2018}. A problem that comes with this approach
is that drawing set-values from independent uniform distributions leads to an
unfavourable distribution of the resulting \glspl{IPF} in the PQ-plane and
extreme points on the margins of the \gls{FOR} are not captured
well~\cite{heleno2015}.

This is where optimization-based methods come into play. The basic idea behind
these methods is not to randomly sample \glspl{IPF} but systematically identify
marginal IPFs by solving a series of \gls{OPF} problems~\cite{silva2018}. In
addition to better coverage of the \gls{FOR}, optimization-based approaches
have the advantage of higher computational efficiency. An important drawback is
however that, except for approaches which solve the \gls{OPF} heuristically,
solving the underlying \gls{OPF} requires an explicit grid model
\cite{contreras2018}. On the other hand, the only heuristic approach published
so far, suffers from poor automatability as it relies on manual tweaking of
hyperparameters \cite{sarstedt2021}. 

In practice, considering the huge size of \glspl{DG}, complete data related to
grid topology (data related to operating equipment incl.\ its characteristics
and topological connections) is not commonly stored \cite{singh2010}, which
complicates parametrization of explicit grid models. 
In such circumstances, black-box machine learning (ML) models trained on
measurement data provided by current smart meters can be an alternative to
physics-based, explicit grid models \cite{barbeiro2014}. 
Due to their compatibility with black-box grid models, we argue that it is
worthwhile to research and improve data-driven approaches to determine the
\gls{FOR} of \glspl{DG}. This paper is intended to point out and analyse the
problem of leptocurtic distribution of \glspl{IPF} with na\"ive sampling
strategies and thus to underpin the need for more effective and more efficient
data-driven sampling strategies, such as those published by
\textcite{contreras2021}, when this paper was already advanced. 

The remainder of the paper is structured as follows: A survey on existing
approaches (data-driven and optimization-based) and the contribution of this
paper are given in \autoref{sec:related_work}.  Next, in
\autoref{sec:experiment_setup} the construction of a series of synthetic
feeders with increasing number of nodes is explained.  On the basis of these
feeders we performed our sampling experiments, the results and evaluation of
which are presented in \autoref{sec:experiment_results}.

Finally, in \autoref{sec:conclusion} the paper is summarized, the conclusion is
given and an idea for a distributed sampling strategy is presented which does
not suffer from unfavourable distribution of resulting \glspl{IPF}.

\section{Survey on grid flexibility aggregation methods and contribution of
this paper}
\label{sec:related_work}
As outlined in the introduction, relevant literature can be grouped into two main
categories: Data-driven/stochastic and optimization-based approaches for exploring the
\gls{FOR} of \glspl{DG}.

\subsection{Data-driven approaches}
\label{sec:data_driven_approaches}

\textcite{heleno2015} are the first to come up with the idea of estimating the
flexibility range in each primary substation node to inform the \gls{TSO} about
the technically feasible aggregated flexibility of \glspl{DG}. In order to
enable the \gls{TSO} to perform a cost/benefit evaluation, they also include
the costs associated with adjusting the originally planned operating point of
flexible resources in their algorithm. In the paper two variants of a Monte
Carlo simulation approach are presented, which differ in the assignment of
set-values to the flexible resources. While in the first approach independent
random set-values for changing active and reactive power injection are
associated to each flexible resource, in the second approach a negative
correlation of one between generation and load at the same bus was considered.
In a direct comparison of the two presented approaches, the approach with
negative correlation between generation and load at the same bus performs
better and results in a wider flexibility range and lower flexibility costs
with a smaller sample size. Nevertheless, even with this approach, the
capability to find marginal points in the \gls{FOR} is limited. Therefore, in
the outlook the authors suggest the formulation of an optimization problem in
order to overcome the limitations of the Monte Carlo simulation approach,
increasing the capability to find extreme points of the \gls{FOR} and reducing
the computational effort. In \textcite{silva2018}, which is discussed in the
next subsection, the authors take up this idea again.

\textcite{mayorgagonzalez2018} extend in their paper the methodology presented
by \textcite{heleno2015}. First, they describe an approach to approximate the
\gls{FOR} of an \gls{ADG} for a particular point in time assuming that all
influencing factors are known. For this, they use the first approach of
\textcite{heleno2015} for sampling \glspl{IPF} (the one that does not consider
correlations).  That is, random control scenarios are generated by assigning
set-values from independent uniform distributions to all controllable units. In
contrast to~\cite{heleno2015}, no cost values are calculated for the resulting
\glspl{IPF}. Instead, for describing the numerically computed \gls{FOR} with
sparse data, the region is approximated with a polygon in the complex plane. In
addition, a probabilistic approach to assess in advance the flexibility
associated to an \gls{ADG} that will be available in a future time interval
under consideration of forecasts which are subject to uncertainty is proposed.
The authors mention that for practical usage the computation time for both
approaches has to be significantly reduced. However, the problem of
unfavourable distribution of the resulting \gls{IPF} point cloud, when drawing
control scenarios from independent uniform distributions, which is a mayor
factor for the low computational efficiency, is not discussed.

When this paper was already advanced, \textcite{contreras2021} came up with new
sampling methods for data-driven approaches. They show that, when focusing the
vertices of the flexibility chart of flexibility providing units during
sampling, the quality of the data-driven approach can be dramatically improved
in comparison to the nai\"ve sampling. On top of that, they present
a comparison of OPF-based and data-driven approaches, whose results show that
with their improved sampling strategies both approaches are capable of
assessing the \gls{FOR} of radial distribution grids but for grids with large
number of buses OPF-based methods are still better suited.

\subsection{Optimization-based approaches}
\label{sec:optimization_based_approaches}
\textcite{silva2018} address the main limitation of their sampling-based
approach in~\textcite{heleno2015}, namely estimating extreme points in the
\gls{FOR}. To this end, they propose a methodology which is based on
formulating an optimization problem with below-mentioned objective function,
whose minimization for different ratios of $\alpha$ and $\beta$ allows to
capture the perimeter of the flexibility area.

\begin{equation}
    \alpha \: P_{\mathit{DSO} \rightarrow \mathit{TSO}}
    + \beta \: Q_{\mathit{DSO} \rightarrow \mathit{TSO}}
\end{equation}

\noindent where $P_{\mathit{DSO} \rightarrow \mathit{TSO}}$ and
$Q_{\mathit{DSO} \rightarrow \mathit{TSO}}$ are the active and reactive power
injections at the TSO-DSO boundary nodes.  \textcite{silva2018} work out that
the underlying optimization problem represents an OPF problem. Due to its
robust characteristics they use the primal-dual, a variant of the interior
point methods to solve it.  The methodology was evaluated in simulation and
validated in real field-tests on MV distribution networks in France. The
comparison of simulation results with the random sampling algorithm
in~\textcite{heleno2015} shows the superiority of the optimization-based
approach by illustrating its capability to identify a larger flexibility area
and to do it within a shorter computing time.

\textcite{capitanescu2018} propose the concept of active-reactive power (PQ)
chart, which characterizes the short-term flexibility capability of active
distribution networks to provide ancillary services to \gls{TSO}. To support
this concept, an AC optimal power flow-based methodology to generate PQ
capability charts of desired granularity is proposed and illustrated in
a modified 34-bus distribution grid.

\textcite{contreras2018} present a linear optimization model for the
aggregation of active and reactive power flexibility of distribution grids at
a \gls{TSO}-\gls{DSO} interconnection point. The power flow equations are
linearized by using the Jacobian matrix of the Newton-Raphson algorithm. The
model is complemented with non-rectangular linear representations of typical
flexibility providing units, increasing the accuracy of the distribution grid
aggregation. The obtained linear programming system allows a considerable
reduction of the required computing time for the process. At the same time, it
maintains the accuracy of the power flow calculations and increases the
stability of the search algorithm while considering large grid models.

\textcite{fortenbacher2020} present a method to compute reduced and aggregated
distribution grid representations that provide an interface in the form of
active and reactive power (PQ) capability areas to improve \gls{TSO}-\gls{DSO}
interactions. Based on a lossless linear power flow approximation they derive
polyhedral sets to determine a reduced PQ operating region capturing all
voltage magnitude and branch power flow constraints of the \gls{DG}. While
approximation errors are reasonable, especially for low voltage grids,
computational complexity is significantly reduced with this method.

\textcite{sarstedt2021} provide a detailed survey on stochastic and
optimization based methods for the determination of the \gls{FOR}. They come up
with a comparison of different \gls{FOR} determination methods with regard to
quality of results and computation time. For their comparison they use the
Cigr\'e medium voltage test system. On top of that, they present a \gls{PSO}
based method for \gls{FOR} determination. 

\subsection*{Contributions of this paper}
In summary, it can be stated that optimization-based approaches show high
computational efficiency with good coverage of the \gls{FOR}. However, methods
used for solving the underlying \gls{OPF} problem rely---except for heuristic
approaches, which have other drawbacks---on explicit grid models of the
\gls{DG}, which must be parametrized with grid topology data often not
available in practice. Data-driven approaches, on the other hand, do not
require explicit grid modeling and are compatible with black-box grid models,
but suffer from low computational efficiency and poor coverage of peripheral
regions of the \gls{FOR}, when using conventional sampling strategies.

This is where our approach comes in. We are heading towards improved sampling
strategies for data-driven approaches, which mitigate the weak points of
data-driven methods (low computational efficiency and poor coverage of
\gls{FOR}) while retaining their advantage of being compatible with black-box
grid models. As a basis for this, in this paper we are the first to come up
with an experiment setup by means of which the problem of resulting convoluted
distribution of \glspl{IPF} with na\"ive sampling strategies can be analyzed
and pointed out in an easily reproducible manner. 

\section{Experiment setup}
\label{sec:experiment_setup}
\begin{figure}
  \includegraphics[width=\linewidth]{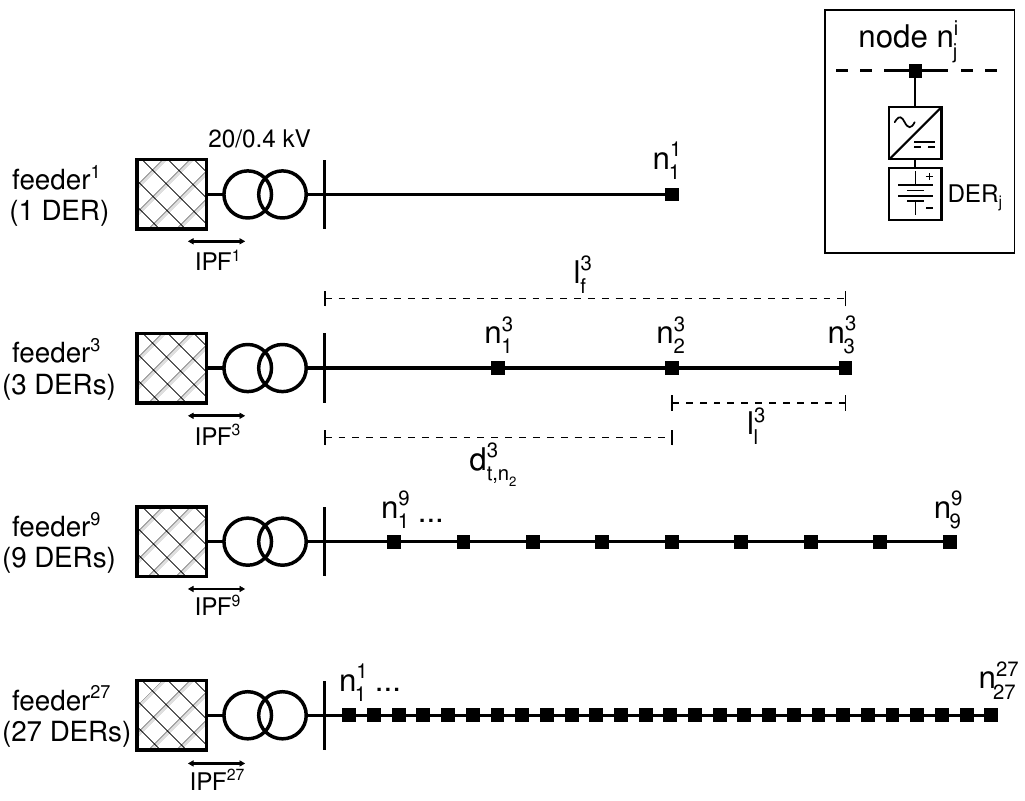}
  \caption{Synthetic \SI{0.4}{kV} feeders}
  \label{fig:grid_sketch}
\end{figure}

To show the shortcomings of na\"ive sampling strategies, we apply a sampling
strategy that draws set-values for the controllable \glspl{DER} from
independent uniform distributions to a series of synthetic \SI{0.4}{kV} feeders
as shown in \cref{fig:grid_sketch}.  The feeders are characterized by an
increasing number of nodes. To be able to consider the effect of the number of
nodes on the \gls{IPF}-sample as isolated as possible, both, the total
installed power and the average transformer-node distance are chosen equal for
all feeders. The installed power is distributed equally among all connected
\glspl{DER}:

\begin{equation}
  \label{eq:installed_power}
  P^i_{\mathit{inst}, \mathit{DER}_j} = \frac{P^i_{\mathit{inst}, \mathit{DERs}}}{N^i},
\end{equation}

\noindent where $P^i_{inst, DER_j}$ is the installed power of the DER connected
to the $j$th node $n^i_j$ of feeder $i$, $P^i_{inst,DERs}$ is the total
installed power of feeder $i$ and $N^i$ is the number of nodes of feeder $i$.
Nodes are equally distributed along feeders as shown in \cref{fig:grid_sketch}
and the line length between adjacent nodes $l^i_l$ of feeder $i$ with length
$l^i_f$ is:

\begin{equation}
  \label{eq:line_length}
  l^i_l = \frac{l^i_f}{N^i}.
\end{equation}

\noindent The transformer-node distance of node $n^i_j$ is:

\begin{equation}
  \label{eq:transformer_node_distance}
  d^i_{t,n_j} = l^i_l \cdot j.
\end{equation}

\noindent With \eqref{eq:transformer_node_distance} the average
transformer-node distance $\overline{d^i_{t,n}}$ of feeder $i$  can be written
as:

\begin{equation}
\label{eq:avrg_transformer_node_distance}
\begin{split}
  \overline{d^i_{t,n}} & = \frac{1}{N^i} \sum_{j=1}^{N^i} d^i_{t,n_j} \\
   & = \frac{1}{N^i} \sum_{j=1}^{N^i} l^i_l \cdot j = \frac{l^i_l}{N^i} \sum_{j=1}^{N^i} j \\
   & = \frac{l^i_l}{N^i} \cdot \frac{N^i_j \cdot (N^i_j + 1)}{2}.
\end{split}
\end{equation}

\noindent Resolved after the line length $l^i_l$, the result is:

\begin{equation}
  \label{eq:line_length_avrg_transformer_node_distance}
  l^i_l = \overline{d^i_{t,n}} \cdot \frac{2}{N^i + 1}.
\end{equation}

\noindent With \eqref{eq:line_length} and
\eqref{eq:line_length_avrg_transformer_node_distance} the length of feeder
$i$ results in:

\begin{equation}
  \label{eq:feeder_length}
  l^i_f = \overline{d^i_{t,n}} \cdot \frac{2 N^i}{N^i + 1}.
\end{equation}

For our experiments we have constructed four synthetic \SI{0.4}{kV}~ feeders.
The feeders differ in the number of nodes $N^i$, which has been set to \num{1},
\num{3}, \num{9} or \num{27} respectively.  Line length $l^i_l$ and feeder
length $l^i_f$ have then been calculated according to
\eqref{eq:line_length_avrg_transformer_node_distance} and
\eqref{eq:feeder_length}. There is one \gls{DER} connected to each node and the
installed power $P^i_{inst, DERs}$ is distributed evenly among the \glspl{DER}
according to \eqref{eq:installed_power}.  To be able to cover the entire
flexibility area of the feeders including its border areas where voltage band
violations and/or line overloadings can be observed, all \glspl{DER} are
inverter-connected battery storages because they offer maximum flexibility with
regard to both, active and reactive power provision. The dimensioning of the
inverters has been chosen in such a way that a power factor $\cos{\phi}$ of 0.9
can be kept, when the maximum active power is delivered:
\begin{equation}
 \label{eq:apparent_power}
 |S|^i_{max,DER_j} = \frac{P^i_{inst, DER_j}}{\cos{\phi}} 
  = \frac{P^i_{inst, DER_j}}{0.9}.
\end{equation} 
Active and reactive power ranges of the battery storages are thus:
\begin{equation}
\thickmuskip=-.1mu
\thinmuskip=-.1mu
\begin{aligned}
 \label{eq:active_power_range}
 \left[ P^i_{min,DER_j}, P^i_{max,DER_j}  \right] & = 
 \left[ -P^i_{inst,DER_j}, P^i_{inst,DER_j} \right] \\
 \left[ Q^i_{min,DER_j}, Q^i_{max, DER_j}  \right] & = 
 \left[ -|S|^i_{max,DER_j}, |S|^i_{max,DER_j} \right].\phantom{\Bigg]} 
\end{aligned}
\end{equation}
Values for the technical parameters of the four feeders including connected
\glspl{DER} are listed in \cref{tab:grids_configuration}. 

\begin{table*}[t]
\caption{Configuration of the synthetic feeders}
\label{tab:grids_configuration}
  \begin{tabu}{CCCCCCCC}
    \toprule
    \# DERs & $\mathrm{P_{inst, DER_j}}$ (kW) & $\mathrm{|S|_{max,DER_j}}$ (kVA)
    & Feeder Length (m) & Line Length (m)
    & Line Type & Voltage Band (pu) & Trafo Type\\
    \midrule
    1 & 200.0 & 222.2 & 400 & 400 & NAYY 4x150 SE & 
    0.9--1.1 & 0.4 MVA 20/0.4~kV \\
    3 & 66.7  & 74.1  & 600 & 200 & NAYY 4x150 SE & 
    0.9--1.1 & 0.4 MVA 20/0.4~kV \\
    9 & 22.2  & 24.7  & 720 & 80  & NAYY 4x150 SE & 
    0.9--1.1 & 0.4 MVA 20/0.4~kV \\
    27 & 7.4  & 8.2   & 771 & 29  & NAYY 4x150 SE & 
    0.9--1.1 & 0.4 MVA 20/0.4~kV \\
    \bottomrule
  \end{tabu}
\end{table*}

For all feeders we conduct the sampling in such a way that for each \gls{DER}
and each sample element we independently draw real and reactive power values
from uniform distributions:
\begin{equation}
\label{eq:uniform_distribution_power_ranges}
\begin{aligned}
\mathcal{X}^i_{P, DER_j} &\sim \mathcal{U}\left[P^i_{min,DER_j},  P^i_{max,DER_j}\right] \\
\mathcal{X}^i_{Q,DER_j} &\sim \mathcal{U}\left[Q^i_{min,DER_j},  Q^i_{max,DER_j}\right].
\end{aligned}
\end{equation}
After assigning active and reactive power values to each \gls{DER}, the
\emph{pandapower} library \cite{thurner2018a} calculates the power flow. This
way we generate a sample of size \num{2500} for each feeder.

Following this, the sample elements are first classified with regard to their
adherence to grid constraints and in case of non-adherence with regard to the
type of grid constraint violation (i.e., voltage band violation, line overload,
or both).  Second, inverter constraints are taken into account and  the sample
elements are classified with regard to adherence to both, grid and inverter
constraints. In this case, sample elements are only classified as feasible, if
neither grid constraints nor device constraints for any of the connected
inverters occur.  In case of non-feasibility we distinguish depending on the
type of constraint violation (i.e.,\ grid constraint violation, inverter
constraint violation, or both).

Finally we plot the classification results in the domain of active and reactive
\glspl{IPF} $P_{IPF}$ and $Q_{IPF}$. 

\section{Experiment results}
\label{sec:experiment_results}
\begin{figure}
\centering
  \includegraphics{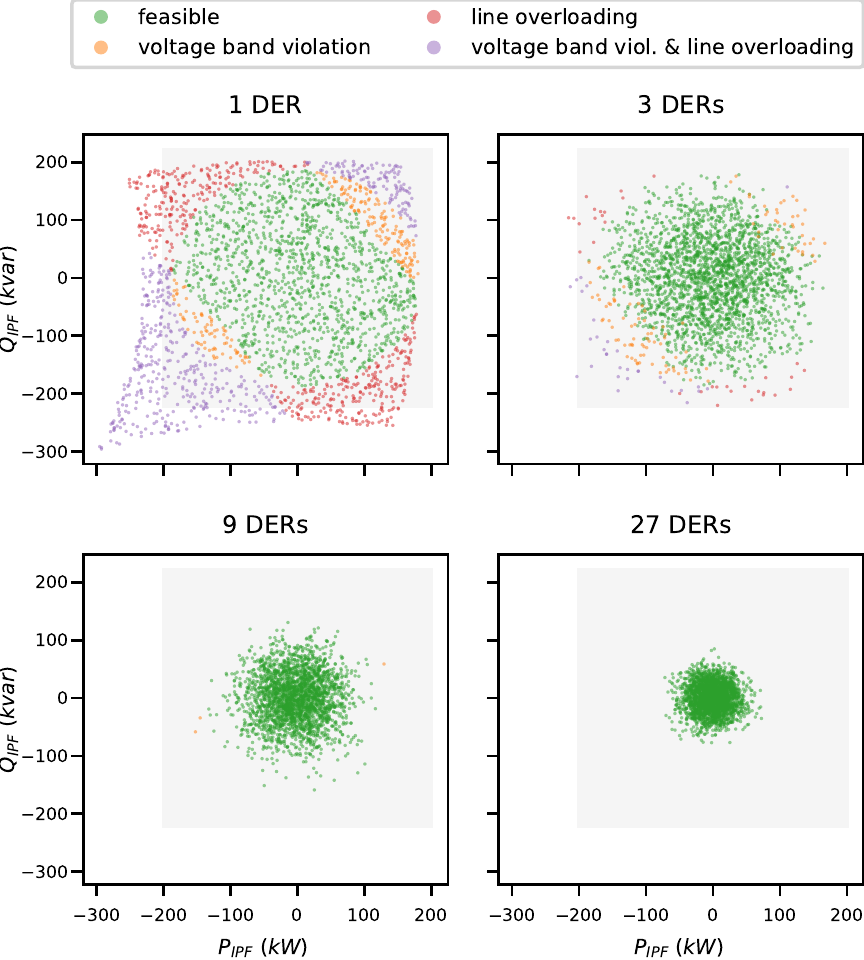}
  \caption{Results of na\"ive sampling strategy classified by feasibility
  with regard to grid constraints (voltage band limits and max. line loading);
  inverter constraints are neglected}
  \label{fig:eval_sampling_grid_only}
\end{figure}

\begin{figure}
\centering
  \includegraphics[]{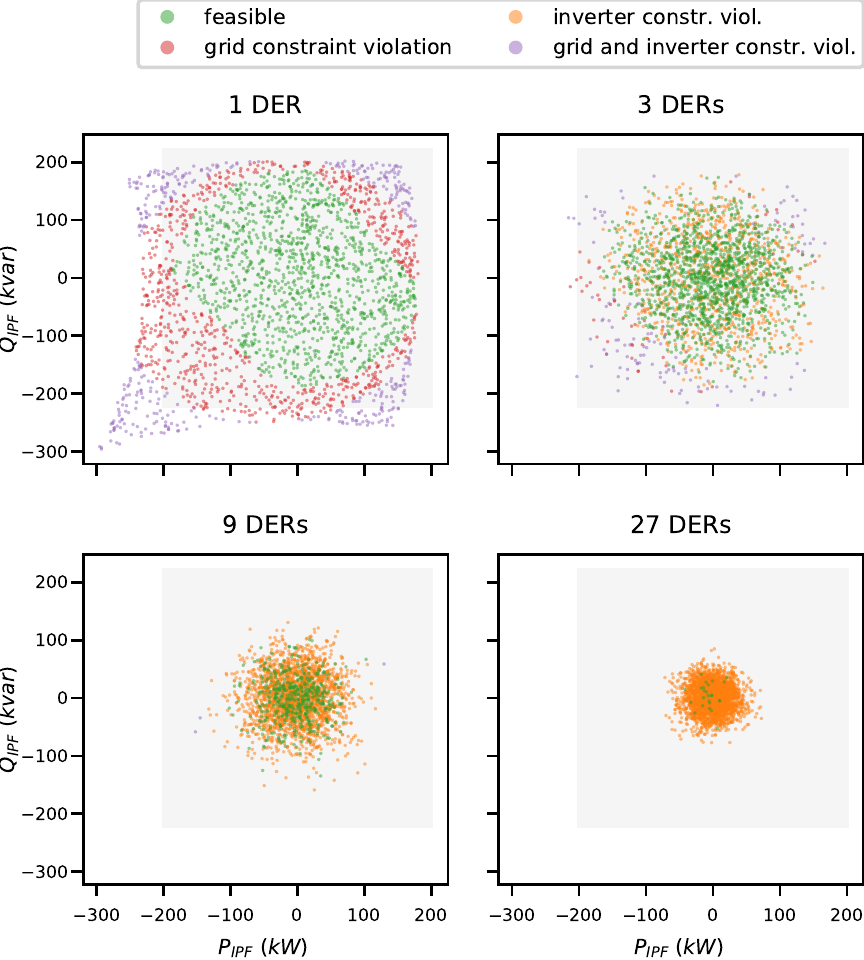}
  \caption{Results of na\"ive sampling strategy classified by feasibility with regard
  to both, grid and inverter constraints; please note: green dots 
  are plotted above orange dots, i.e.~no green dots are covered by orange dots}
  \label{fig:eval_sampling_grid_inverter}
\end{figure}
The resulting plots are shown in \cref{fig:eval_sampling_grid_only}
and~\ref{fig:eval_sampling_grid_inverter}.  The sampling has been performed
once for each feeder from \cref{fig:grid_sketch}.  This means that
\cref{fig:eval_sampling_grid_only} and~\ref{fig:eval_sampling_grid_inverter}
only differ in how the sample elements are classified. While for
\cref{fig:eval_sampling_grid_only} only grid constraints have been considered,
\cref{fig:eval_sampling_grid_inverter} also incorporates inverter constraints.
Both figures consist of four subplots---one for each of the four feeders from
\cref{fig:grid_sketch}.  Each dot of the point clouds represents one sample
element---so every subplot contains \num{2500} dots.  The shaded grey area
marks the theoretically known aggregated power limit for the \glspl{DER}:
\begin{equation}
\thickmuskip=-.1mu
\begin{aligned}
 \label{eq:active_power_range_feeder}
 \left[ P^i_{min,DERs}, P^i_{max,DERs}  \right] & = 
 \left[ -P^i_{inst,DERs}, P^i_{inst,DERs} \right] \\
 \left[ Q^i_{min,DERs}, Q^i_{max, DERs}  \right] & = 
 \left[ -|S|^i_{max,DERs}, |S|^i_{max,DERs} \right].\phantom{\Big]}
\end{aligned}
\end{equation} 

For the 1~\gls{DER} case, shape and structure of the point cloud look as one
would expect from the configuration.  It largely covers the grey area---only
slightly skewed and shifted towards lower active and reactive power values,
which results from active and reactive power consumption of grid elements
(lines and transformer).  However, when increasing the number of \glspl{DER},
the \emph{convolution problem} becomes obvious. With 3~\glspl{DER}, the
feasible area is still covered to some extent, but the point density already
decreases strongly towards the edges. In case of 9 \glspl{DER} the point
density in the edges has decreased to such an extent that hardly any sample
elements are detected which show grid constraint violations. Finally, with
27~\glspl{DER} the point cloud has collapsed to a fraction of the grey area and
only a small part of the theoretical \gls{FOR} is covered.

From the 3~\glspl{DER} subplot in \cref{fig:eval_sampling_grid_only} it can be
seen that with increasing number of \glspl{DER} not only the region covered by
the sample collapses, but at the same time the border between feasible and
infeasible elements (with regard to grid constraints) becomes less distinct:
The absence of a sharp border between feasible and non-feasible \glspl{IPF}
complicates the use of multi-class classification for identifying the \gls{FOR}
from the sample and indicates the use of a one-class classifier for that
purpose. 

\cref{fig:eval_sampling_grid_inverter}, which additionally considers inverter
constraints, shows an other problem of the nai\"ve sampling approach: In this
consideration, not only the total area covered by the sample decreases, but
also the share of feasible examples shrinks sharply, such that with
27~\glspl{DER} only very few sample elements are identified which violate
neither grid nor inverter constraints. 

This is because with the nai\"ve sampling approach power values are assigned to
each \gls{DER} at once. After that, the power flow calculation is performed and
only at the very end the feasiblity with regard to grid and inverter
constraints is checked. Even if the constraints of only a single inverter are
violated, the example is classified as non-feasible with regard to inverter
constraints. If, for example, for a single converter one third of the possible
power setpoints violate constraints, the likelihood to observe no constraint
violations with \(N\) inverters amounts to \[\left(1 - \frac{1}{3}\right)^N.\]
For $N=27$ inverters this would amount to approximately \num{1.76e-5}.

One way to address this would be to perform a successive sampling as proposed
by \textcite{bremer2013a} for the use case of active power planning.  With
successive sampling the evaluation of inverter constraints is done immediately
after the assignment of setpoints to single \glspl{DER} and in case of
non-feasibility drawing of setpoints is repeated until a valid configuration is
found.  The power flow calculation would then be carried out only after
setpoints compatible with inverter constraints have been found for each
\gls{DER}. \todo{Zitat Bremer einfuegen!} 

\begin{figure}
  \includegraphics{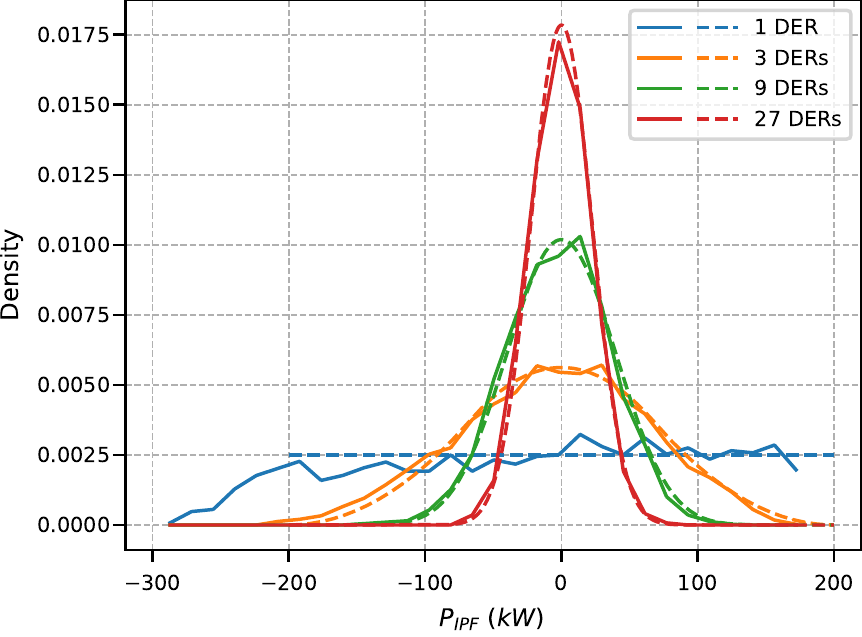}
  \caption{Frequency density of active IPFs resulting from our experiments (solid lines) compared with probability density 
  function of Bates distribution on the interval $[-P^i_{inst,DERs}, P^i_{inst,DERs}]$ (dashed lines)}
  \label{fig:epdf}
\end{figure}

To illustrate the \emph{convolution problem}, in \cref{fig:epdf} we plot the
frequency density of active \glspl{IPF} resulting from our sampling against the
\gls{PDF} of the Bates distribution.  The Bates distribution is the continuous
probability distribution of the mean of \(n\)~independent uniformly distributed
random variables on the unit interval and thus closely related to the
Irwin-Hall distribution, which describes the sum of \(n\)~independent uniformly
distributed random variables. More general, in statistics the probability
distribution of the sum of two or more independent random variables is the
convolution of their individual distributions.  For the variant of the Bates
distribution generalized to arbitrary intervals $\left[a,b\right]$: 
\begin{equation}
 \label{eq:bates_distribution}
\mathcal{X}_{(a,b)} = \frac{1}{n}\sum_{k=1}^n \mathcal{U}_k(a,b)
\end{equation} 
this results in the following equation defining the \gls{PDF}:
\begin{equation}
 \label{eq:pdf_bates_distribution}
f(x) = 
\begin{cases}
\begin{aligned}
\sum_{k=0}^{n} \Bigg[ &(-1)^k \binom{n}{k} 
  \left( \frac{x-a}{b-a}-\frac{k}{n} \right)^{n-1} \\
&\sgn\left( \frac{x-a}{b-a}-\frac{k}{n} \right) \Bigg]
\end{aligned}& \text{if } x\in[a,b] \\
0 & \text{otherwise}.
\end{cases}
\end{equation} 

Comparison with the Bates distribution is motivated by the fact that with
equations (\ref{eq:installed_power}), (\ref{eq:active_power_range}) and
(\ref{eq:uniform_distribution_power_ranges}) the active power range from which
we draw values during the sampling can be written as follows:
\begin{equation}
\label{eq:distribution_p_setpoints_single_der}
\begin{aligned}
\mathcal{X}^i_{P,DER_j} 
&\sim \mathcal{U}^i_j \left[\frac{-P^i_{inst,DERs}}{N^i},  
\frac{P^i_{inst,DERs}}{N^i}\right] \\
&= \frac{1}{N^i} \mathcal{U}^i_j\left[-P^i_{inst,DERs}, P^i_{inst,DERs}\right].
\end{aligned}
\end{equation}
For a single sample element the active \gls{IPF} $P^i_{IPF}$ is made up of the
sum of active power injections of connected \glspl{DER} $P^i_{DER_{j}}$ and the
grid losses $P^i_{loss}$:
\begin{equation}
\label{eq:ipf_components}
    P^i_{IPF} = \sum_{j=1}^{N_i} P^i_{DER_{j}} + P^i_{loss}
\end{equation}
We are interested in the distribution $\mathcal{X}^i_{P,IPF}$ of active
\glspl{IPF}. Ignoring grid losses $P^i_{loss}$ in (\ref{eq:ipf_components}),
with equations (\ref{eq:distribution_p_setpoints_single_der}) and
(\ref{eq:ipf_components}) we can write:
\begin{equation}
\label{eq:distribution_ipfs}
\begin{aligned}
\mathcal{X}^i_{P,IPF}
&\sim \sum_{j=1}^{N_i} \frac{1}{N^i} \mathcal{U}^i_j\left[-P^i_{inst,DERs},
P^i_{inst,DERs}\right] \\
&= \frac{1}{N^i} \sum_{j=1}^{N_i} \mathcal{U}^i_j\left[-P^i_{inst,DERs},
P^i_{inst,DERs}\right],
\end{aligned}
\end{equation}
which is exactly the Bates distribution on the interval $\left[-P^i_{inst,DERs},
P^i_{inst,DERs}\right]$. 

\section{Conclusion and Future Work}
\label{sec:conclusion}
Aggregating the flexibility potential of \glspl{DG} is an important
prerequisite for effective \gls{TSO}-\gls{DSO} coordination in electric power
systems with high share of generation located in the \gls{DG} level. In this
paper we first gave an overview of existing flexibility aggregation methods and
categorized them in terms of whether they are data-driven/stochastic or
optimization-based.  Following this, we discussed the strengths and weaknesses
of both approaches (stochastic and optimization-based) and motivated the
investigation of improved sampling strategies for data-driven approaches. As
a basis for this, we presented an experimental setup by means of which we
demonstrated and analyzed the shortcomings of nai\"ve sampling strategies with
focus on the problem of resulting leptokurtic distribution of \glspl{IPF}.

In future work we will investigate approaches for mitigating the
\emph{convolution problem}.  One idea is to formulate the sampling as
a distributed optimization problem whose objective function takes into account
the uniformity of the resulting set of \glspl{IPF}. First experiments in this
direction with the \gls{COHDA} protocol by \textcite{hinrichs2017} and with
Ripleys-K as metric for the evaluation of the distribution show promising
results. 

Additionally, we are working on making \gls{OPF}-based methods compatible with
black-box grid models by solving the underlying \gls{OPF} with the help of
evolutionary algorithms such as the covariance matrix adaptation evolution
strategy (CMA-ES) \cite{hansen2006} or REvol, an algorithm which was originally
developed for training artificial neural networks \cite{veith2014}.
Furthermore, we want to investigate if the total number of required objective
function evaluations can be reduced when sampling the border of the FOR in one
run by dynamically adapting the underlying objective function.

\section*{Acknowledgements}
This work was funded by the Deutsche Forschungsgemeinschaft
(DFG, German Research Foundation) -- 359921210.

\printbibliography
\end{document}